\documentclass[aps,prl,reprint,groupedaddress]{revtex4-1}
\usepackage{graphicx} % Include figure files
\usepackage{dcolumn} % Align table columns on decimal point
\usepackage{bm}% bold math
\usepackage{hyperref}% add hypertext capabilities

\begin{document}

\title{ Production of neutron-rich heavy nuclei around N = 162 in multinucleon transfer reactions }
\author{Cheng Peng}
\author{Zhao-Qing Feng}
\email{Corresponding author: fengzhq@scut.edu.cn}

\affiliation{School of Physics and Optoelectronics, South China University of Technology, Guangzhou 510640, China}

\date{\today}

\begin{abstract}

Within the framework of the dinuclear system model, the production mechanism of neutron-rich heavy nuclei around N = 162 has been investigated systematically. The isotopic yields in the multinucleon transfer reaction of $^{238}$U + $^{248}$Cm was analyzed and compared the available experimental data. Systematics on the production of superheavy nuclei via $^{238}$U on $^{252,254}$Cf, $^{254}$Es and $^{257}$Fm is investigated. It is found that the shell effect is of importance in the formation of neutron-rich nuclei around N=162 owing to the enhancement of fission barrier. The fragments in the multinucleon transfer reactions manifest the broad isotopic distribution and are dependent on the beam energy. The polar angles of the fragments tend to the forward emission with increasing the beam energy. The production cross sections of new isotopes are estimated and heavier targets are available for the neutron-rich superheavy nucleus formation. The optimal system and beam energy are proposed for the future experimental measurements.

\begin{description}
\item[PACS number(s)]
25.70.Hi, 25.70.Lm, 24.60.-k      
\end{description}
\end{abstract}

\maketitle

\section{I. Introduction}

The nucleus is a quantum manybody system composed of protons and neutrons. The fission limits the amount of nuclear mass \cite{Bo,Ho}. The shell effect increases the fission barrier and enable the survival of superheavy nucleus (SHN). Exploring the limit of element or atomic nucleus is a long-term task, especially close to the 'island of stability' predicted by the shell model \cite{WD,AF,Mo94}. Up to now, there are more than 3000 nuclides existing on the nuclide chart, and most of them are located in the proton-rich region \cite{Th16}. For the past few decades, new transactinide nuclei have been synthesized by the fusion-evaporation reactions in different laboratories. Before 1974, the isotopes of element Z=93-106 were created by the neutron capture reactions and heavy-ion fusion reactions. Combinations with a doubly magic nucleus or nearly magic nucleus are usually chosen owing to the larger reaction $Q$ values. Reactions with $^{208}$Pb or $^{209}$Bi based targets were firstly proposed by Oganessian et al. \cite{Og75}. The SHEs from Bh to Cn were synthesized in the cold fusion reactions at GSI (Darmstadt, Germany) with the heavy-ion accelerator UNILAC and the SHIP separator \cite{Ho00,Mu15}. Experiments on the synthesis of element Nh (Z=113) in the $^{70}$Zn+$^{209}$Bi reaction have been performed successfully at RIKEN (Tokyo, Japan) \cite{Mo04}. However, it is very difficulty to create the superheavy isotopes beyond Nh in the cold fusion reactions because of very low cross section ($\sigma<$0.1 pb). The superheavy elements from Fl (Z=114) to Og (Z=118) have been synthesized at the Flerov Laboratory of Nuclear Reactions (FLNR) in Dubna (Russia) with the double magic nuclide $^{48}$Ca bombarding actinide nuclei \cite{Og99,Og06,Og15}, in which more neutron-rich SHN were produced and identified by the subsequent $\alpha$-decay chain. With constructing the new facilities in the world such as RIBF (RIKEN, Japan), SPIRAL2 (GANIL in Caen, France), FRIB (MSU, USA), HIAF (IMP, China), the SHNs on the 'island of stability' by using the neutron-rich radioactive beams induced fusion reactions or via the multinucleon transfer (MNT) reactions might be possible in experiments.

The MNT reactions and deep inelastic heavy-ion collisions were extensively investigated in experiments since 1970s, in which the new neutron-rich isotopes of light nuclei and proton-rich actinide nuclei were observed \cite{Art71,Art73,Art74,Hil77,Gla79,Moo86, Wel87}. The reaction mechanism and fragment formation were investigated thoroughly, i.e., the energy and angular momentum dissipation, two-body kinematics, shell effect, fission of actinide nuclei etc. Recently, the MNT reactions have been extensively investigated in experiments for new isotope production. The MNT reactions have the advantage of extensive isotope distributions, e.g., more than 100 nuclides with Z=82-100 in the reaction of $^{48}$Ca + $^{248}$Cm and five new neutron-deficient isotopes $^{216}$U, $^{219}$Np, $^{223}$Am, $^{229}$Am and $^{233}$Bk \cite{De15}. The reaction mechanism is concentrated on the neutron-rich nuclei around the shell closure N = 126. The MNT fragments were measured in the reactions of $^{136}$Xe + $^{208}$Pb \cite{JW15,EE12}, $^{136}$Xe + $^{198}$Pt \cite{YX15}, $^{156,160}$Gd + $^{186}$W \cite{EM17}, and $^{238}$U + $^{232}$Th \cite{SK18}. The neutron-rich nuclides around N = 126 have important application in understanding the origin of heavy elements from iron to uranium in the \emph{r}-process of astrophysics. It has been confirmed that the shell closure plays an important role on the production of neutron-rich nuclei and more advantage with the multinucleon transfer (MNT) reactions in comparison to the projectile fragmentation \cite{Ku14}.

Several models have been proposed for describing the MNT reactions, i.e., the dinuclear system (DNS) model \cite{Fe09,GG10,GG81}, the GRAZING model \cite{Win94,Win95}, the dynamical model based on multidimensional Langevin equations \cite{Zag15}. Moreover, the microscopic approaches based on the nucleon degree of freedom, the time-dependent Hartree-Fock (TDHF) approach \cite{CC09,KK13,WN18,Gu19} and the improved quantum molecular dynamics (ImQMD) \cite{TJ08,ZK13,ZK15} are also used to describe the MNT reactions. Some interesting issues have been investigated, e.g., the production cross sections of new isotopes, total kinetic energy spectra and polar angle distribution of transfer fragments, structure effect on the fragment formation etc. There are still some open problems for the MNT reactions, i.e., the mechanism of preequilibrium cluster emission, the stiffness of nuclear surface during the nucleon transfer process, the mass limit of new isotopes with stable heavy target nuclides, interplay of nuclear fission and particle evaporation in the MNT dynamics, etc. The extremely neutron-rich beams are favorable for creating neutron-rich heavy or superheavy nuclei owing to the isospin equilibrium \cite{Das94}.

In this work, the systematics on the production of neutron-rich heavy nuclei around N = 162 has been investigated. The article is organized as follows. In  Section II we give a brief description of the DNS model. In Section III, the transfer cross section and angular distribution of actinide collisions are studied. Summary and perspective on the neutron-rich heavy nuclei are shown in Section IV.

\section{II. Brief description of the model}
\subsection{ The MNT fragment formation }

The DNS concept was assumed to be formal at the touching configuration in nucleon collision and was proposed by Volkov for describing the deep inelastic transfer reaction\cite{Vo20}. The sticking of projectile and target forms DNS. Sticking times are on the order of $(10^{-21}-10^{-20})$s. Once DNS is formed,the exchange of nucleons and energy stars. The motion of nucleons in the interacting potential is governed by the single-particle Hamiltonian \cite{Fe07}. Only the particles within the valence space are active for nucleon excitation and transfer\cite{W24,S25}. The transition probability is related to the local excitation energy and nucleon transfer\cite{Fe07}\cite{Fe06}, which is microscopically derived from the interaction potential in valence space. The local excitation energy is determined by the dissipation energy from the relative motion and the potential energy surface of the DNS. The dissipation of the relative motion and angular momentum of the DNS is described by the Fokker-Planck equation \cite{Fe07a}.  The cross sections of the primary fragments (Z$_{1}$, N$_{1}$) after the DNS reaches the relaxation balance are calculated as follows:
\begin{eqnarray}
	\sigma_{pr}(Z_{1},N_{1},E_{c.m.})=&&\sum^{J_{max}}_{J=0}\sigma_{cap}(E_{c.m.},J)\int f(B) \nonumber\\
	&&  \times P(Z_{1},N_{1},E_{1},J_{1},B)dB
\end{eqnarray}

The cross sections of secondary fragments that is the product of the primary fragment after the evaporation of particles in the MNT reactions are evaluated by
\begin{eqnarray}
	\sigma_{sur}(Z_{1},N_{1},E_{c.m.})=&&\sum^{J_{max}}_{J=0}\sigma_{cap}(E_{c.m.},J)\int f(B) \nonumber\\
	&&  \times P(Z_{1},N_{1},E_{1},J_{1},B) \nonumber \\
	&&  \times W_{sur}(E_{1},J_{1},s)dB
\end{eqnarray}
Here, $E_{1}$ and $J_{1}$ denote the excitation energy and the angular momentum for the fragment$(Z_{1},N_{1})$ with the attempting barrier B, respectively. The maximal angular momentum $J_{max}$ is taken to be the grazing collision of two nuclei. J is the relative angular momentum. The expression of capture cross is $\sigma_{cap}=\pi\hbar^{2}(2J+1)/(2\mu E_{c.m.})T(E_{c.m.},J)$ and the cross is calculated within the Hill-Wheeler formula and the barrier distribution approach. The transferred cross section is smoothed with the barrier distribution and the function is taken to have the Gaussian form $f(B) =\frac{1}{N}exp[-((B-B_{m})/\Delta)^{2}]$, with the normalization constant satisfying the unity relation $\int f(B)dB=1$. The quantities $B_{m}$ and $\Delta$ are evaluated by $B_{m}=(B_{C} + B_{S})/2$ and $\Delta = (B_{C} - B_{S})/2$, respectively. $B_{C}$ and $B_{S}$ are the Coulomb barrier at waist-to-waist orientation and the minimum barrier from varying the quadrupole deformation parameters of the colliding partners.

The nucleon transfer is related to the relative motion by solving a set of microscopically derived master equations by distinguishing protons and neutrons \cite{Fe06,Fe07}. The time evolution of the distribution probability $P(Z_{1},N_{1},E_{1},t)$ for a DNS fragment 1 with proton number $Z_{1}$ and neutron number $N_{1}$ and excitation energy $E_{1}$ is governed by the master equations as follows,
\begin{eqnarray}
	&&\frac{d P(Z_{1},N_{1},E_{1}, t)}{dt}           \nonumber\\
	&=& \sum_{Z^{'}_{1}}W_{Z_{1},N_{1};Z^{'}_{1},N_{1}}(t)[d_{Z_{1},N_{1}}P(Z^{'}_{1},N_{1},E^{'}_{1},t)  \nonumber\\
	&& -d_{Z^{'}_{1},N_{1}}P(Z_{1},N_{1},E_{1},t)]     \nonumber\\
	&& +\sum_{N^{'}_{1}}W_{Z_{1},N_{1};Z_{1},N^{'}_{1}}(t)\times[d_{Z_{1},N_{1}}P(Z_{1},N^{'}_{1},E^{'}_{1},t)  \nonumber\\
	&& -d_{Z_{1},N^{'}_{1}}P(Z_{1},N_{1},E_{1}, t)]    \nonumber\\
\end{eqnarray}
Here the $W_{Z_{1},N_{1};Z_{1}^{\prime},N_{1}}$ ($W_{Z_{1},N_{1};Z_{1},N_{1}^{\prime}}$) is the mean transition probability from the channel $(Z_{1},N_{1},E_{1})$ to $(Z_{1}^{\prime},N_{1},E_{1}^{\prime})$ (or $(Z_{1},N_{1},E_{1})$ to $(Z_{1},N_{1}^{\prime},E_{1}^{\prime})$), and $d_{Z_{1},N_{1}}$
denotes the microscopic dimension corresponding to the macroscopic state $(Z_{1},N_{1},E_{1})$. Sequential one-nucleon transfer is considered in the model with the relation of $Z_{1}^{\prime}=Z_{1}\pm 1$ and $N_{1}^{\prime }=N_{1}\pm 1$.

DNS is affected by potential energy surface (PES) during nuclear transfer. The projectile and target is transferred under the action of the driving potential. The potential energy surface of the DNS is given by
\begin{eqnarray}
	U(\{\alpha\})=&&B(Z_{1},N_{1})+B(Z_{2},N_{2})\nonumber \\&&-\left[B(Z,N)+V^{CN}_{rot}(J)\right]
	+V(\{\alpha\})
\end{eqnarray}
Here Z and N are the proton and neutron number of the composite system respectively with $Z_{1}+Z_{2}=Z$ and $N_{1}+N_{2}=N$ \cite{Fe09}.  The symbol $\{\alpha\}$ denotes the sign of the quantities $Z_{1},N_{1}, Z_{2}, N_{2}; J, R; \beta_{1}, \beta_{2}, \theta_{1}, \theta_{2}$. The $B(Z_{i},N_{i}) (i=1,2)$ and $B(Z,N)$ are the negative binding energies of the fragment $(Z_{i},N_{i})$ and the compound nucleus $(Z,N)$, respectively. The $V^{CS}_{rot}$ is the rotation energy of the compound system. The $\beta_{i}$ represent the quadrupole deformations of the two fragments at ground state. The $\theta_{i}$ denote the angles between the collision orientations and the symmetry axes of deformed nuclei. The interaction potential between fragment $(Z_{1},N_{1})$ and $(Z_{2},N_{2})$ includes the nuclear, Coulomb and centrifugal parts. In the calculation, the distance $R$ between the centers of the two fragments is chosen to be the value at the touching configuration, in which the DNS is assumed to be formed. So the PES depends on the proton and neutron numbers of the fragments.

The survival probability $W_{sur}$ of each fragment is evaluated with a statistical approach based on the Weisskopf evaporation theory\cite{P22}, in which the excited primary fragments are cooled by evaporating $\gamma$ rays, light particles including neutrons, protons, $\alpha$ in competition with binary fission. Shell correction, odd-even effect, Q value, etc. are the structure effects that could be particularly significant in the formation of the primary fragments and in the decay process. the probability in the channel of evaporating the x-th neutron, the y-th proton and the z-alpha is expressed as
\begin{eqnarray}
    W_{sur}(E^{*}_{CN},x,y,z,J)= && P(E^{*}_{CN},x,y,z,J)          \nonumber\\
	&&  \times \prod_{i=1}^{x} \frac{\Gamma_{n} (E^{*}_{i},J)} {\Gamma_{tot}(E^{*}_{i},J)} \prod_{j=1}^{y}\frac{\Gamma_{p}(E^{*}_{j},J)}{\Gamma_{tot}(E^{*}_{j},J)}       \nonumber \\
	&&  \times \prod_{k=1}^{z}\frac{\Gamma_{\alpha}(E^{*}_{k},J)}{\Gamma_{tot}(E^{*}_{k},J)}
\end{eqnarray}

Here the $E^{*}_{CN}$, J are the excitation energy evaluated from the mass table\cite{PM95} and the spin of the excited nucleus, respectively. The total width $\Gamma_{tot}$ is the sum of partial widths of particle evaporation, $\gamma$ -emission and fission. The excitation energy $E^{*}_{s}$ before evaporating the s-th particle is evaluated by
\begin{eqnarray}
E^{*}_{s+1}=E^{*}_{s}-B^{n}_{i}-B^{p}_{j}-B^{\alpha}_{k}-2T_{s}
\end{eqnarray}
with the initial condition  $E^{*}_{1}$ =  $E^{*}_{CN}$ and s = i+j+k. The $B^{n}_{i}$, $B^{p}_{j}$, $B^{\alpha}_{k}$ are the separation energy of the i-th neutron, j-th proton, k-th alpha, respectively. The nuclear temperature $T_{i}$ is given by $E^{\ast}_{i}=a T_{i}^{2} - T_{i}$ and the level density parameter $a$. The evaporation width and realization probability are described in detail in reference \cite{P22}.

\subsection{  Angular distribution }
The detection of emission angle of fragments formed in MNT reaction of primary fragments is beneficial to the placement of detectors in experiments. Different systems and incident energies affect the angular distribution. In this work, we use a deflection function method to evaluate the fragment angle which can be changed by the mass of fragment, angular momentum, and incident energy. The deflection angle is composed of the Coulomb and nuclear interactions as \cite{Ch17,GW78}
\begin{eqnarray}
\Theta(l_{i}) = \Theta(l_{i})_{C}+\Theta(l_{i})_{N}
\end{eqnarray}
with
\begin{eqnarray}
	\Theta(l_{i})_{C}=2\arctan \frac{Z_{1}Z_{2}e^{2}}{2E_{inc}b}.
\end{eqnarray}
The Coulomb deflection function is given by the Rutherford function. The expression of nuclear deflection angle is shown as
\begin{eqnarray}
\Theta(l_{i})_{N} = -\beta\Theta(l_{i})_{C}^{gr}\frac{l_{i}}{l_{gr}}(\frac{\delta}{\beta})^{l_{i}/l_{gr}}.
\end{eqnarray}
Here $\Theta(l_{i})_{C}^{gr}$ is the Coulomb scattering angle at the grazing angular momentum $l_{gr}$ and $l_{gr}=0.22R_{int}[A_{red}(E_{c.m.}-V(R_{int}))]^{1/2}$. The $l_{i}$ is the incident angular momentum.  $A_{red}$, and $V(R_{int})$ are reduced mass of projectile and target and interaction potential, respectively. $R_{int}$ is the distance from the entrance channel. The $\delta$ and $\beta$ are parameterized by fitting the deep inelastic scattering in massive collisions as
\begin{eqnarray}
\label{eq10}
\beta = &&  75 f(\eta) + 15,   \qquad  \eta <375      \nonumber       \\
&& 36 \exp(-2.17\times 10 ^{-3} \eta),   \qquad  \eta \geq 375
\end{eqnarray}

\begin{eqnarray}
\label{eq11}
\delta = && 0.07 f(\eta) + 0.11,     \qquad  \eta <375        \nonumber   \\
&&  0.117 \exp(-1.34\times 10 ^{-4} \eta),  \qquad  \eta \geq 375
\end{eqnarray}
and
\begin{equation}
\label{eq13}
f(\eta) = [1 + \exp{\frac{\eta-235}{32}}]^{-1}
\end{equation}
where the Sommerfeld parameter $\eta = \frac{Z_1Z_2e^2}{\upsilon}$, and the relative velocity $\upsilon = \sqrt{\frac{2}{A_{red}}(E_c.m. - V(R_{int}))}$. For the $i-$th DNS fragment, the emission angle is determined by $\Theta_{i}(l_{i})=\Theta(l_{i})\xi_{i}/(\xi_{1}+\xi_{2})$ with the moment of inertia $xi$.

\section{III. Results and discussion}

The synthesis of superheavy nuclei in the fusion-evaporation reactions has entered the bottleneck stage because of the lack of neutron number of formed compound nucleus. There are several pathways to reach the 'island of stability', i.e., the MNT reactions, incomplete fusion reactions and complete fusion reactions induced with radioactive nuclides etc. The damped collisions of two actinide nuclei were investigated and motivated for producing superheavy nuclei in 1970s at Gesellschaft f\"{u}r Schwerionenforschung (GSI) \cite{Hil77,Gla79}. The experimental data were reanalyzed for investigating the MNT dynamics, in particular for the transactinide production \cite{Kr13}. As a test of the DNS model, the available data in collisions of $^{238}$U+$^{248}$Cm at the beam energy of E$_{lab}$ = 7.0 MeV/nucleon are compared as shown in Fig. 1. It is obvious that the isotopic yields are consistent with the available data. The maximal yield of each isotopic chain is located on the line of $\beta-$stability. The distribution structure of the isotopic yields is related to the PES, nucleon transfer dynamics, nuclear fission, particle evaporation etc. The extension to the neutron-rich side is possible but monotonically decreasing with the mass number of target-like fragment.

%%%%%%%%%%%%%%%%%%%%%%%%%%%%%%%%%%%% figure 1 %%%%%%%%%%%%%%%%%%%%%%
\begin{figure*}
\includegraphics[width=15 cm]{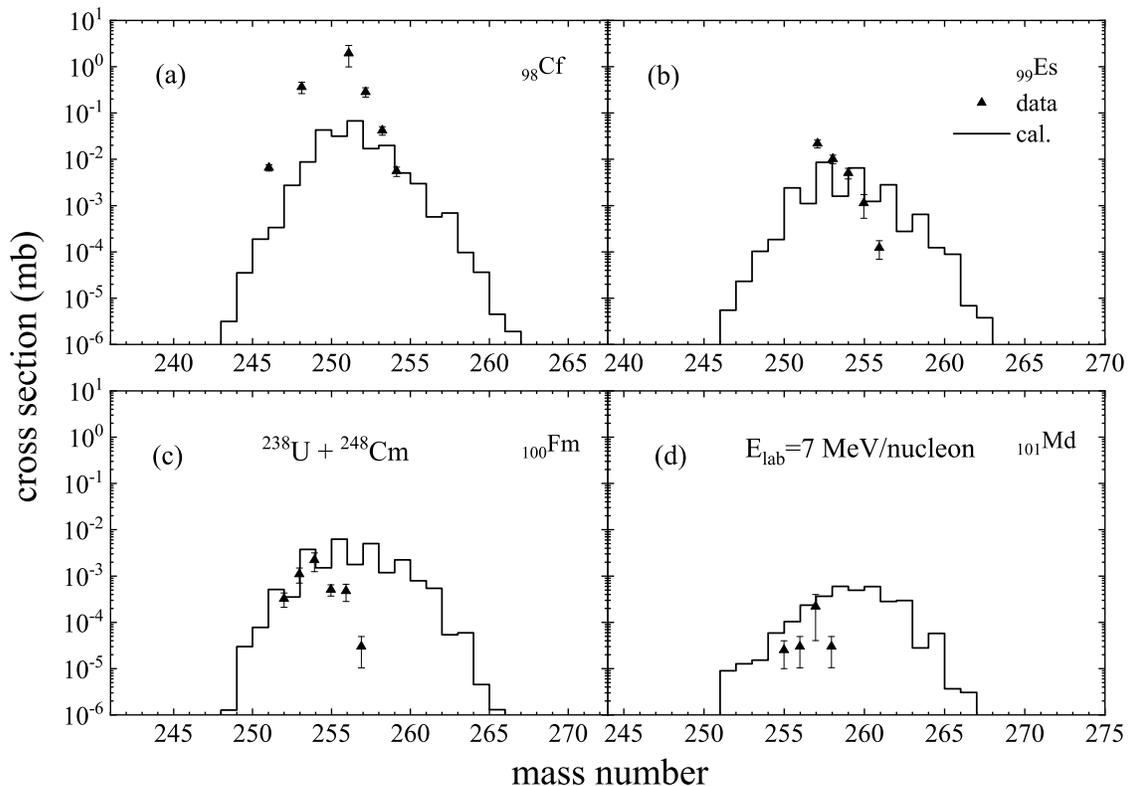}
\caption{Production cross sections of transcurium isotopes in the $^{238}$U+$^{248}$Cm reaction at E$_{lab}$ =7.0 MeV/nucleon and compared with the available experimental data at GSI with error bars \cite{Kr13}. }
\end{figure*}
%%%%%%%%%%%%%%%%%%%%%%%%%%%%%%%%%%%%%%%%%%%%%%%%%%%%%%%%%%%%%%%

The properties of neutron-rich nuclei beyond N=126 are particular of importance in understanding the nucleosynthesis in stellar evolution, the nuclear fission dynamics, the new decay modes of isotopes on the 'island of stability' etc. The shape evolution, shell effect, nucleon transfer dynamics, dissipation mechanism of the relative motion and rotation of the dinuclear system influence the MNT fragment formation. Shown in Fig. 2 is a comparison of isotopic distribution in the $^{238}$U induced reactions on $^{252}$Cf, $^{254}$Es and $^{257}$Fm at the beam energy of 7 MeV/nucleon. The open symbols denote the new isotopes in the nuclear chart \cite{Wa21}.  Compared to the neutron-deficient nuclei via the fusion-evaporation reactions, the MNT reactions exhibit a broad isotopic distribution. The production cross sections decrease with increasing the proton number of fragments. The heavier target $^{257}$Fm is favorable for creating the neutron-rich isotopes owing to the larger isospin ratio neutron/proton of colliding system. However, the different systems almost have the same order magnitudes of isotopic yields in the proton-rich domain. The maximal yields of the isotopic spectra in the MNT reactions correspond to the line of $\beta-$stability, in which the shell effect and fission barrier are of significance on the spectrum structure.

%%%%%%%%%%%%%%%%%%%%%%%%%%%%%%%%%%%% figure 2 %%%%%%%%%%%%%%%%%%%%%%
\begin{figure*}
\includegraphics[width=15 cm]{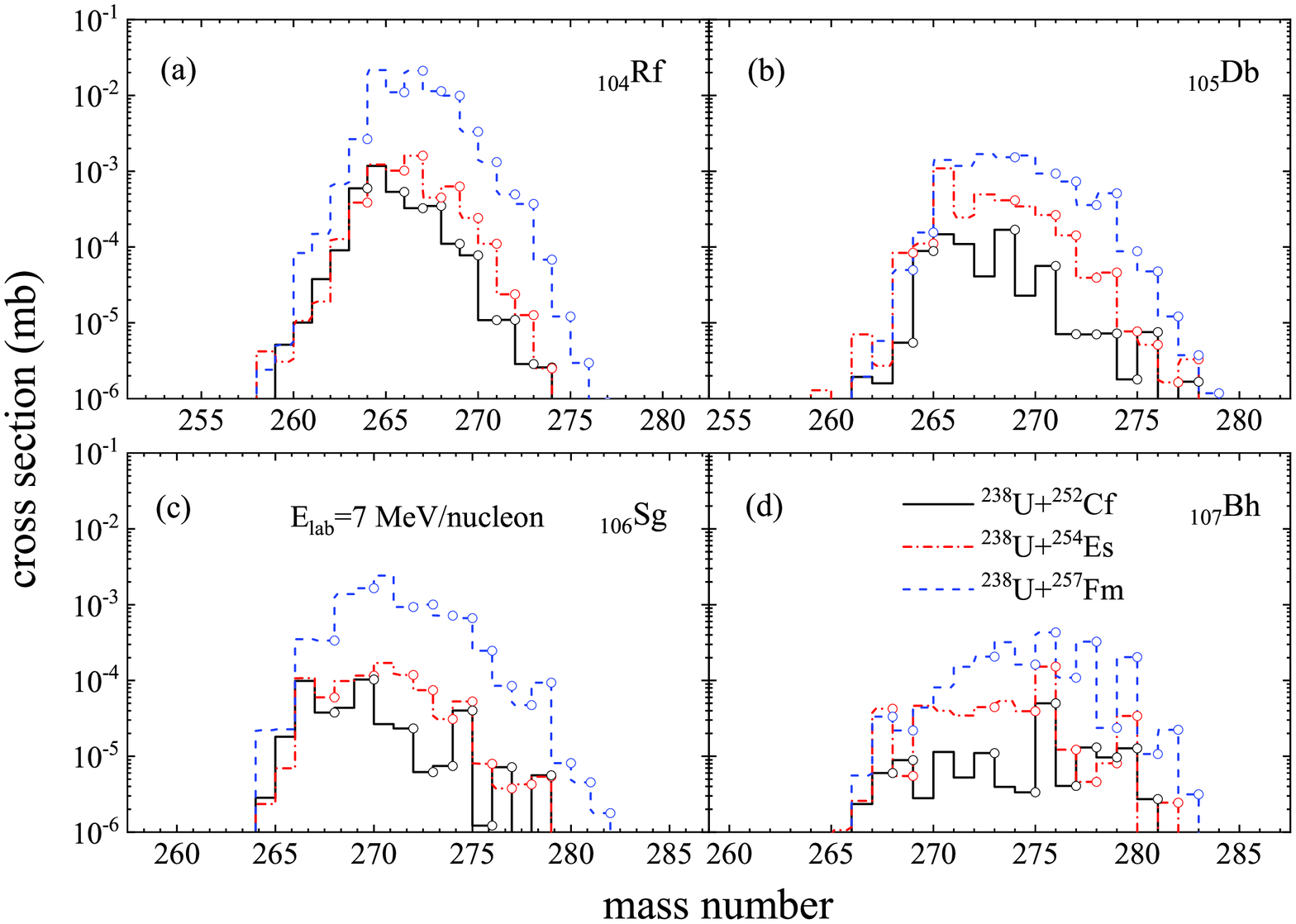}
\caption{Isotopic distribution of Rf, Db, Sg and Bh in the MNT reactions at E$_{lab}$ = 7.0 MeV/nucleon with $^{238}$U bombarding $^{252}$Cf, $^{254}$Es and $^{257}$Fm, respectively. The open symbols denote the new isotopes \cite{Wa21}. }
\end{figure*}
%%%%%%%%%%%%%%%%%%%%%%%%%%%%%%%%%%%%%%%%%%%%%%%%%%%%%%%%%%%%%%%

The isotopic dependence of initial systems in the MNT reactions is of significance in the selection of projectile-target combination in experiments and the investigation of isospin dissipation mechanism in theories. Figure 3 shows the isotopic distribution of Rf, Db, Sg and Bh in the reactions of $^{238}$U + $^{252}$Cf (soild line) and $^{238}$U + $^{254}$Cf (dotted line) at E$_{lab}$ =7.0 MeV/nucleon. The isotopic targets are selected for producing the MNT fragments. It is obvious that the overall products move to the neutron-rich side with the neutron-rich target $^{254}$Cf. However, the isotopes in the proton-rich region exhibit an opposite trend. The entrance system with the larger neutron/proton ratio is available for producing the neutron-rich fragments owing to the isospin equilibrium in the dissipation process. The shell effect and isospin relaxation play significant roles on the MNT fragment formation \cite{Fe17,NF17,NF18}.

%%%%%%%%%%%%%%%%%%%%%%%%%%%%%%%%%%%% figure 3 %%%%%%%%%%%%%%%%%%%%%%
\begin{figure*}
	\includegraphics[width=15 cm]{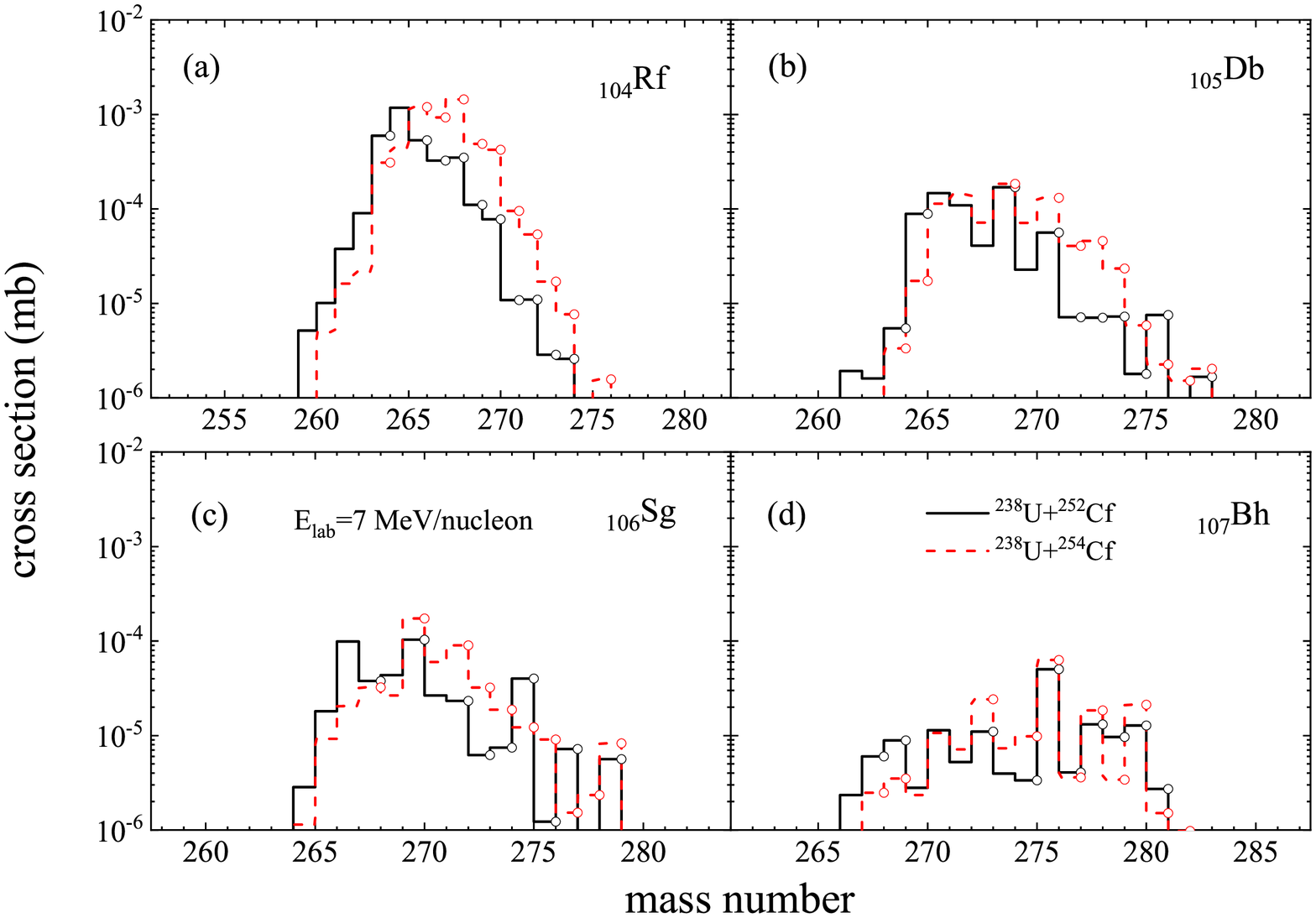}
	\caption{Comparison of Rf, Db, Sg and Bh production in the MNT reactions at E = $_{lab}$7.0 MeV/nucleon with $^{238}$U bombarding $^{252}$Cf and $^{254}$Cf, respectively. The open symbols denote the new isotopes with the recent nuclear mass evaluation \cite{Wa21}.}
\end{figure*}
%%%%%%%%%%%%%%%%%%%%%%%%%%%%%%%%%%%%%%%%%%%%%%%%%%%%%%%%%%%%%%%

The beam energy influences the dissipation energy into the DNS and consequently the MNT fragment formation. The more local excitation energy is obtained with increasing the incident energy and leads to the wider domain nucleon transfer \cite{NF17}. A number of rare isotopes might be created with more energetic nuclear collisions. However, the less survival of the primary fragment is obtained by the transfer dynamics. The competition of the diffusion of nucleon transfer and the survival of formed fragment leads the isotopic structure of final MNT fragments, which is associated with the beam energy and colliding system. It is found that the total mass and charge distributions of the secondary fragments around the shell closure N = 126 weakly depend on the bombarding energy in the MNT reaction of $^{136}$Xe + $^{198}$Pt \cite{Fe17}. Systematic investigation of energy dependence on the fragment formation in the MNT reactions would be helpful for selecting the optimal beam energy in experiment. Figure 4 shows isotopic distribution of elements of Rf, Db, Sg and Bh in the reaction of  $^{238}$U + $^{252}$Cf at the collision energies E$_{lab}$ = 6.7, 7.0 and 7.4 MeV/nucleon, respectively. The production cross section decreases with the beam energy because of the reduction of survival probability of formed MNT fragment. We calculated the excitation functions of selected nuclides $^{266}$Rf, $^{267}$Db and $^{268}$Sg produced in the reaction of $^{238}$U + $^{252}$Cf as shown in Fig. 5. There exists an optimal incident energy for producing the MNT fragments with the larger cross sections. More beam energy is not suitable for the heavy fragment formation owing to the reduction of survival probability. Roughly, the selection of beam energy around the Coulomb barrier (interaction potential at the touching position) is enough for the MNT reaction.

%%%%%%%%%%%%%%%%%%%%%%%%%%%%%%%%%%%% figure 4 %%%%%%%%%%%%%%%%%%%%%%
\begin{figure*}
	\includegraphics[width=14 cm]{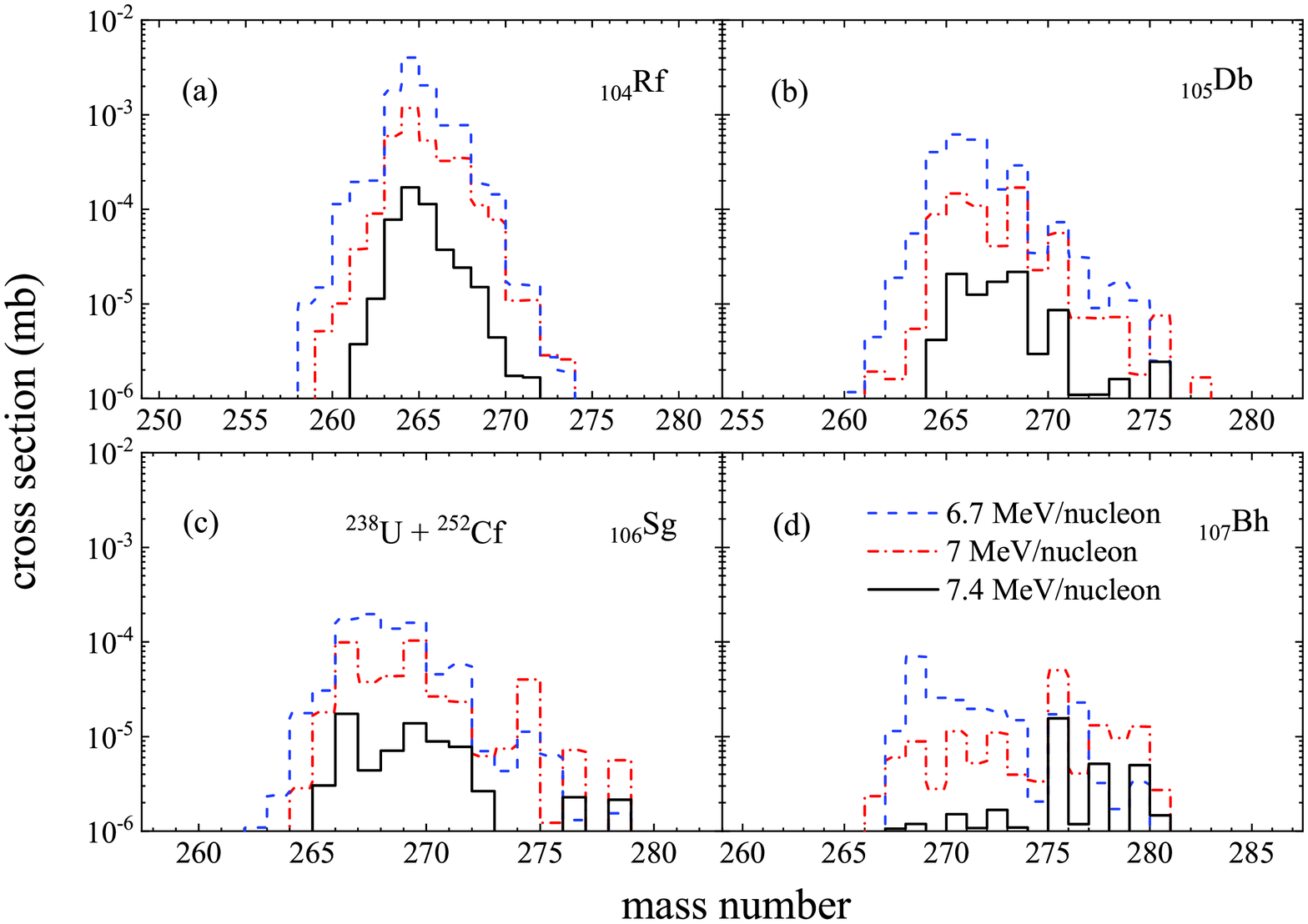}
	\caption{Incident energy dependance of isotopic distribution of Rf, Db, Sg and Bh in the MNT reaction of  $^{238}$U + $^{252}$Cf. }
\end{figure*}
%%%%%%%%%%%%%%%%%%%%%%%%%%%%%%%%%%%%%%%%%%%%%%%%%%%%%%%%%%%%%%%

%%%%%%%%%%%%%%%%%%%%%%%%%%%%%%%%%%%% figure 5 %%%%%%%%%%%%%%%%%%%%%%
\begin{figure*}
	\includegraphics[width=10 cm]{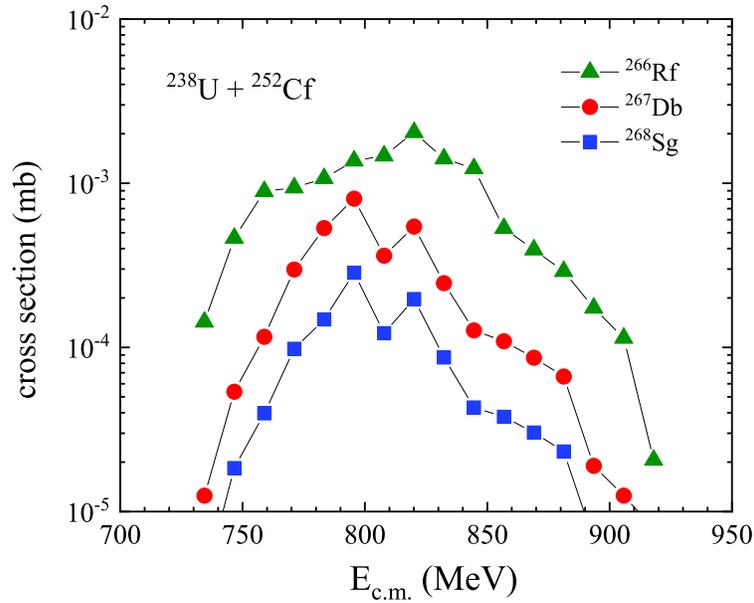}
	\caption{Excitation functions of $^{266}$Rf, $^{267}$Db and $^{268}$Sg production in the MNT reaction of  $^{238}$U + $^{252}$Cf.}
\end{figure*}
%%%%%%%%%%%%%%%%%%%%%%%%%%%%%%%%%%%%%%%%%%%%%%%%%%%%%%%%%%%%%%%

%%%%%%%%%%%%%%%%%%%%%%%%%%%%%%%%%%%% figure 6 %%%%%%%%%%%%%%%%%%%%%%
\begin{figure*}
	\includegraphics[width=10 cm]{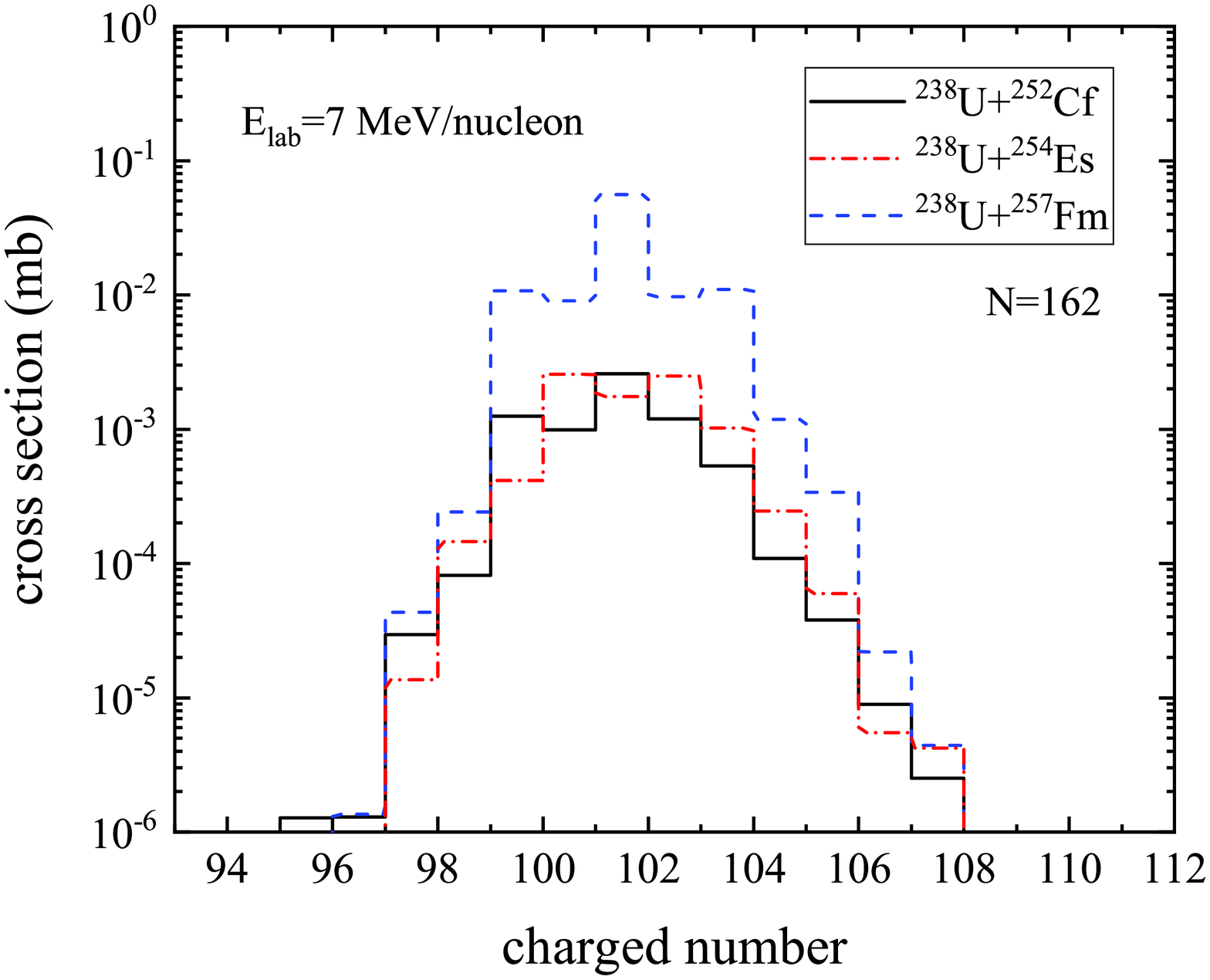}
	\caption{Isotopic fragments with the neutron shell closure N=162 in the MNT reactions with $^{238}$U bombarding $^{252}$Cf, $^{254}$Es and $^{257}$Fm at E$_{lab}$ = 7.0 MeV/nucleon.}
\end{figure*}
%%%%%%%%%%%%%%%%%%%%%%%%%%%%%%%%%%%%%%%%%%%%%%%%%%%%%%%%%%%%%%%

The neutron shell closure around N=126 is favorable for enhancing the cross sections of neutron-rich isotopes and extensive investigation in the MNT reactions with the DNS model \cite{Fe17,Ch20}. The shell effect enhances the fission barrier and enlarges the separation energy of the MNT fragment. On the other hand, the strong shell effect leads to the appearance of pocket in the PES, which also enhances the fragment formation in the nucleon transfer process. A number of neutron-rich isotopes around N = 162 might be created with the damped collisions of two actinide nuclei. More pronounced structure effect is shown in Fig. 6 from the isotonic yields of N=162 in the MNT reactions of $^{238}$U+$^{252}$Cf, $^{238}$U + $^{254}$Es and $^{238}$U+$^{257}$Fm at E$_{lab}$=7.0 MeV/nucleon. The collisions of $^{238}$U+$^{257}$Fm have the larger cross sections because of the neutron/proton ratio and primary fragment yields. The three systems have the same maximal yields around the new isotope $^{263}$Md. The systematic investigation of the new isotope production around N=162 is complicated both in experiments and in theories. Different with the neutron shell closure of N=126, the fission barrier and decay mode is not known in the superheavy region. There are uncertainties of some physical quantities and level spectra in the domain of isotopes around N=162. More experiments and dynamical models in the MNT reactions around N=162 are expected.

%%%%%%%%%%%%%%%%%%%%%%%%%%%%%%%%%%%% figure 7 %%%%%%%%%%%%%%%%%%%%%%
\begin{figure*}
	\includegraphics[width=14 cm]{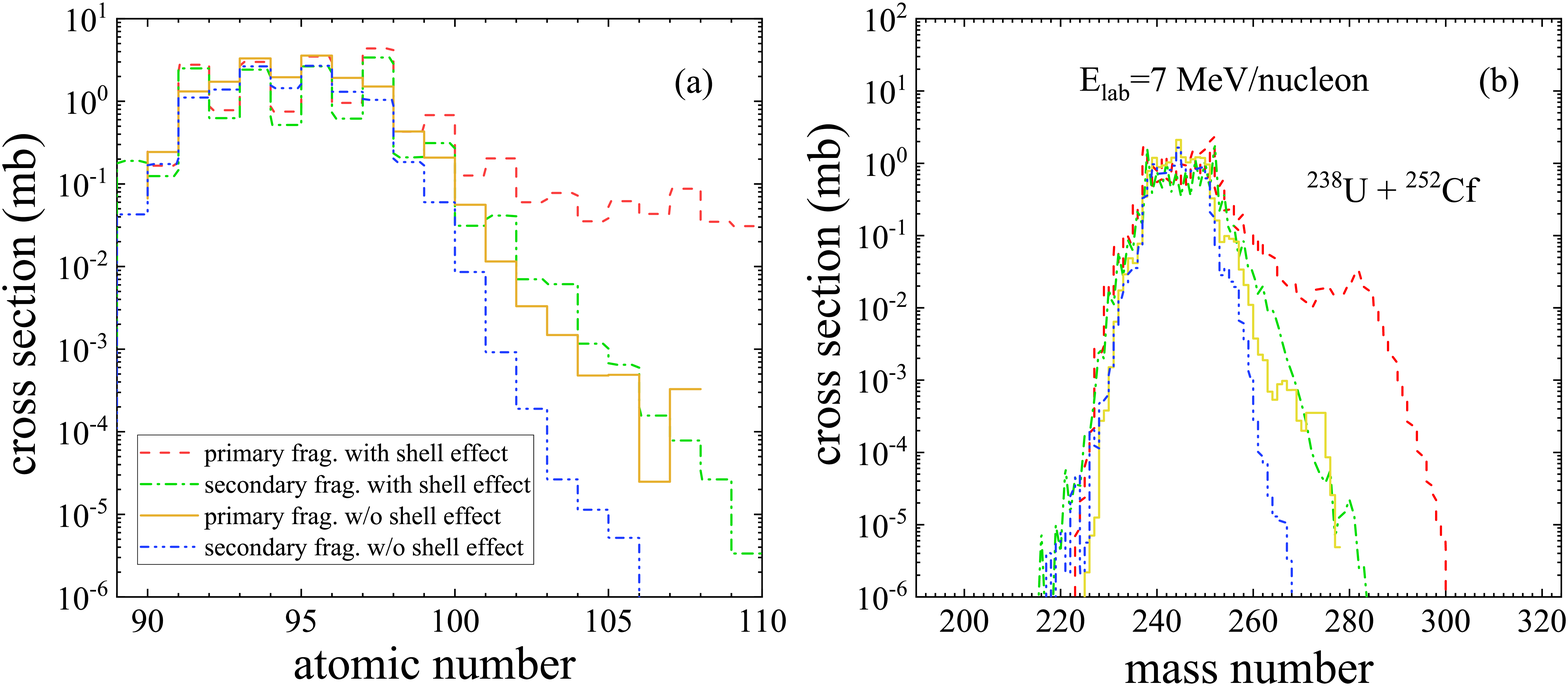}
	\caption{Cross sections as functions of the atomic numbers and the mass numbers of primary and secondary fragments with shell effect or not in the reaction of  $^{238}$U + $^{252}$Cf at E$_{lab}$ = 7.0 MeV/nucleon.}
\end{figure*}
%%%%%%%%%%%%%%%%%%%%%%%%%%%%%%%%%%%%%%%%%%%%%%%%%%%%%%%%%%%%%%%

The collisions of two actinide nuclides provide the possibility to produce the new-neutron isotopes of transfermium elements. The yields of neutron-rich and long-living SHNs in the MNT reactions might be significantly enhanced owing to the shell effect via the 'inverse quasi-fission' process. The shell effect is of significance in the MNT fragment formation, which is manifested via the PES, fission barrier and particle evaporation in the DNS model. Figure 7 shows the production cross sections as functions of atomic number (left panel) and mass number (right panel) in the reaction of  $^{238}$U + $^{252}$Cf at E$_{lab}$ = 7.0 MeV/nucleon. It is obvious that the production cross sections of primary fragments are enhanced around the neutron shell closure N=162, which are caused from the shell effect in the reaction dynamics. The production of secondary fragments is reduced by several orders of magnitude owing to the deexcitation process, in particular in the superheavy region. The decay stage of the MNT fragments is associated with the local excitation energy and angular momentum at the equilibrium time, fission barrier, separation energy of evaporated particles (neutron, proton, deuteron, triton, alpha etc). The strong shell effect can enhance the fission barrier and separation energy, which is favorable the survival of the primary fragment. For the cases without the shell effect, the production cross sections of the primary and secondary fragments monotonically decrease with the mass or charge number, which reduces the formation of heavy rare isotopes in the MNT reactions. The shell effect increases the MNT yields above the two order magnitude. The results are consistent with the model of multidimensional Langevin equations \cite{Zag15}. Inclusion of the structure factors such as the shell effect, odd-even effect, nucleus stiffness etc is particulary significant in the dynamical models for the MNT reactions.

%%%%%%%%%%%%%%%%%%%%%%%%%%%%%%%%%%%% figure 8 %%%%%%%%%%%%%%%%%%%%%%
\begin{figure*}
	\includegraphics[width=14 cm]{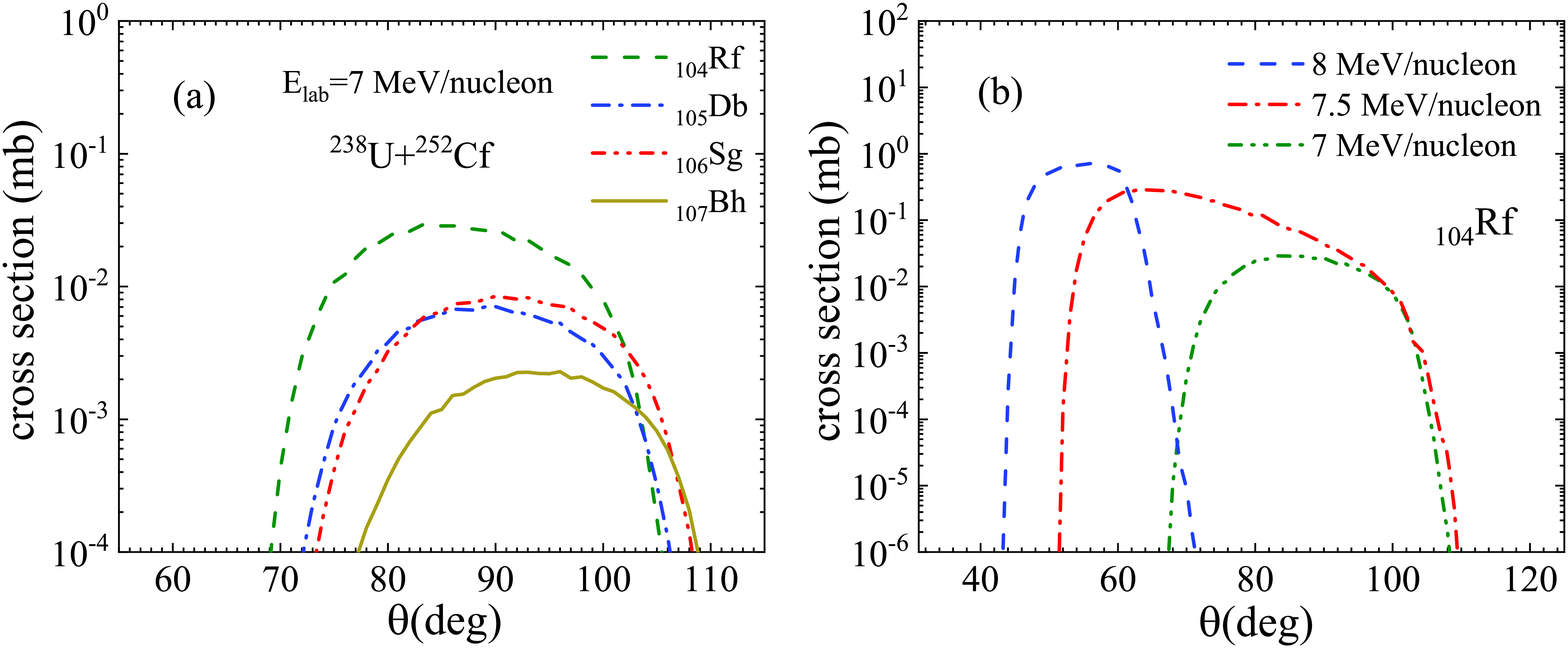}
	\caption{The angular distributions of superheavy nucleus of Z = 104-107 in the laboratory frame in the $^{238}$U+$^{252}$Cf reaction at E$_{lab}$ = 7.0 MeV/nucleon (left panels) and the angular distributions of Rf with different incident energy (right panels). }.
\end{figure*}
%%%%%%%%%%%%%%%%%%%%%%%%%%%%%%%%%%%%%%%%%%%%%%%%%%%%%%%%%%%%%%%

The emission of the MNT fragments is anisotropic and related to the reaction system and beam energy \cite{EE12}. Accurate estimation of the angular distribution would be helpful for managing the detector system in experiments, which is associated with the dynamical characteristics in MNT reactions, i.e., the deformation of the DNS fragments, dissipation of relative energy and angular momentum, nucleon transfer etc\cite{NF20}. Figure 8 shows the selected nuclides with Z=104-107 produced in the MNT reaction of $^{238}$U+$^{252}$Cf at E$_{lab}$ =7.0 MeV/nucleon (left panel) and the angular distribution of Rf at the incident energies of 7, 7.5, 8 MeV/nucleon, respectively (right panel). Slight movement of the emission polar angles with the maximal yields for the SHEs Rf, Db, Sg and Bh appears from 80$^{o}$ to 95$^{o}$ at the incident energy of 7 MeV/nucleon. The nuclides produced in the MNT reactions move to the forward emission with the maximal cross section from 85$^{o}$ at 7 MeV/nucleon to 50$^{o}$ at 8 MeV/nucleon and manifest the narrow angular distribution with increasing the beam energy.

\section{IV. Conclusions}

In summary, the production of neutron-rich isotopes via the MNT reactions was investigated within the DNS model in the $^{238}$U induced reaction on  $^{252,254}$Cf, $^{254}$Es and $^{257}$Fm at the incident energy around Coulomb barrier. By comparing the available experimental data, the DNS model is favorable for describing the MNT reaction dynamics. The shell effect, fission barrier, neutron separation energy and nucleon transfer dynamics influence the formation of MNT fragments. A broad isotopic spectrum is exhibited in the MNT reactions. A number of neutron-rich isotopes might be created around the Coulomb barrier energy. The energy dependence of the neutron-rich nuclide production is weak. But the proton-rich isotopes are related the incident energy. The heavier target is favorable for the SHN production. The yields of MNT fragments monotonically decrease with the mass number. The shell effect, in particular around N=162, results in the appearance of the bump structure in the MNT mass spectra. The phenomena is reduced by the secondary decay process. The MNT fragments tend to be emitted in the forward direction with increasing the incident energy. The slight difference on the angular distribution exists for different isotopes.

\section{Acknowledgements}

This work was supported by the National Natural Science Foundation of China (Projects No. 11722546 and No. 11675226) and the Talent Program of South China University of Technology.

\end{document}